\documentclass[aps,showpacs,twocolumn]{revtex4}

\usepackage{graphicx}
\usepackage{latexsym}
\usepackage{amsmath,amssymb}        
\usepackage[draft=false]{hyperref}

\begin{document}


\title{The three dynamical fates of Boson Stars}


\author{F. S. Guzm\'an}
\affiliation{Instituto de F\'{\i}sica y Matem\'{a}ticas, Universidad
              Michoacana de San Nicol\'as de Hidalgo. Edificio C-3, Cd.
              Universitaria, 58040 Morelia, Michoac\'{a}n,
              M\'{e}xico.}




\begin{abstract}
In this manuscript the three types of late-time behavior of spherically symmetric Boson Stars are
presented, these are: stable configurations, unstable bounded that collapse to form black holes and 
unstable unbounded that explode. The results are found by solving the spherically symmetric Einstein-Klein-Gordon system of equations for a complex scalar field with initial conditions corresponding to a Boson Star. The solution is based on a constrained evolution that uses the method of lines for the scalar field and solves the constraint equations for the geometry on the fly.
\end{abstract}




\maketitle



\section{Introduction}
\label{sec:introduction}

Boson stars (BSs) are self-gravitating systems that help illustrating how 
more complicated systems -like neutron stars- evolve. 
Nevertheless, BSs
have been studied not only as toy models but have also been 
considered as potentially existing astrophysical objects. In this 
sense, BSs can be assumed to exist because they can be 
considered to represent the final stage of zero temperature 
self-gravitating Bose Condensates \cite{GuzmanUrena2006}, 
with regular geometry and smooth matter distribution, with 
no horizons or singularities. Because these objects do not emit in the 
electromagnetic spectrum and do not interact in any way except through gravity, they are transparent. 
In fact BSs are studied as potential black
hole candidates, with results that discard their existence like
those presented in \cite{rees}, others showing that there must be
significant differences when the model considers an accretion
disk \cite{diego-acc}, and a third position showing that by 
choosing the correct boson star, such star can mimic the behavior
of a black hole \cite{Guzman2006}. In a different context, the Newtonian version of BSs has 
been considered as model of galactic halos explaining the galaxy 
formation process under the scalar field dark matter hypothesis 
\cite{GuzmanUrena2006,BernalGuzman2006a,BernalGuzman2006b,mielke_1}.
For reviews on BSs see \cite{jetzer,topical-review}.

The association of BSs with astrophysical objects
may become important when possible astrophysical measurable signals
related to these objects might be at hand soon. This is the case
of gravitational wave (GW) astronomy. There are already some results related
to the GW signals sourced by boson stars, either single stars perturbed 
with shells of particles \cite{Balakrishna2006} or binary boson star
systems \cite{Lehner2007}.
The key point is that as gravitational wave sources are analyzed numerically, the
technology is also applicable to the study of BS systems. 

BS solutions, as happens with other type of general relativistic self-gravitating 
configurations, define stable and unstable branches.  Nevertheless,
the BS unstable branch shows a rather strange property, that is, there
are unstable configurations that collapse into black holes and 
others that disperse away, because the binding energy is allowed to be
negative or positive. It calls the attention that the later configurations have 
not been studied sufficiently, and here we present important properties of this type
of configurations. The results presented in this paper confirm and illustrate, in the full non-linear regime, the three behaviors using the integration of an initial value problem using numerical in spherical symmetry. This study was presented before for the bounded cases \cite{SeidelSuen1990,Balakrishna1998}. An unbounded case was presented  in \cite{Guzman2004} in full 3D numerical simulations and confirmed here with a different set up, which  supports the predictions from perturbation theory (see e.g. \cite{gleiser,Catastrophe}).

The paper is organized as follows: 
in Section \ref{sec:BSs} the usual algorithm to construct BS 
configurations is shown, in Section 
\ref{sec:evolution} the algorithm used here to evolve BSs is presented;
in Section \ref{sec:initialdata} the configurations for the different types of behavior 
are selected; in Section \ref{sec:test1} basic tests are 
shown for stable configurations; in Section \ref{sec:test2} and \ref{sec:test3}
the expected behavior of unstable bounded and unbounded configurations 
(respectively) is presented; finally in Section \ref{sec:remarks} a few comments
are mentioned.


\section{Boson Star configurations}
\label{sec:BSs}

BSs arise from the Lagrangian density of a complex scalar field minimally coupled to 
gravity, that is:

\begin{equation}
{\cal L} = -\frac{R}{\kappa_0} + g^{\mu \nu}\partial_{\mu} \phi^{*}
\partial_{\nu}\phi + V(|\phi|^2),
\label{eq:lagrangian}
\end{equation}

\noindent where $\kappa_0 = 16 \pi G$ in units where $c=\hbar =1$, $\phi$ is the 
scalar 
field, the star 
stands for complex conjugate and $V$ is the potential of 
the field \cite{jetzer,topical-review}. Notice that this Lagrangian density
is invariant under the global $U(1)$ group, and the associated
conserved charge is the amount called number of particles (defined below). When 
the action is varied with respect to the metric 
Einstein's equations arise $G_{\mu\nu} = \kappa_0 T_{\mu\nu}$, with the 
following stress-energy tensor 

\begin{equation}
T_{\mu \nu} = \frac{1}{2}[\partial_{\mu} \phi^{*} \partial_{\nu}\phi +
\partial_{\mu} \phi \partial_{\nu}\phi^{*}] -\frac{1}{2}g_{\mu \nu}
[\phi^{*,\alpha} \phi_{,\alpha} + V(|\phi|^2)].
\label{eq:set}
\end{equation}

\noindent Boson stars are related to the potential 
$V=m^2 |\phi|^2 + \frac{\lambda}{2}|\phi|^4$, although the 
name Boson Star has been applied to solutions using other types of 
potentials (see e.g. \cite{diego-f}). The quantity $m$ is 
understood as the mass of the boson and $\lambda$ is the coefficient of a 
two body self-interaction mean field approximation. The Bianchi identity 
reduces to the Klein-Gordon equation 

\begin{equation}
\left(
\Box - \frac{dV}{d|\phi|^2}
\right)\phi = 0,
\end{equation}

\noindent where $\Box 
\phi=\frac{1}{\sqrt{-g}}\partial_{\mu}[\sqrt{-g}
g^{\mu\nu}\partial_{\nu}\phi]$.

Boson stars (BSs) are spherically symmetric solutions to the above set of equations under a 
particular condition: the scalar field has a harmonic time 
dependence $\phi(r,t) = \phi_0(r) e^{-i \omega t}$, where $r$ is the
radial spherical coordinate. This condition implies that the stress 
energy tensor in (\ref{eq:set}) is time-independent, which implies 
through Einstein's equations that the geometry is also time-independent. 
That is, there is a time-dependent scalar field oscillating upon a 
time-independent geometry whose source is the scalar field itself. 
It is possible to construct solutions for boson stars assuming that 
the metric can be written in normal coordinates as 

\begin{equation}
ds^2=-\alpha(r)^2dt^2 + a(r)^2dr^2 + r^2 d\Omega^2. 
\label{eq:spherical_metric}
\end{equation}

\noindent For these coordinates the Einstein-Klein-Gordon (EKG) system of 
equations reads:

\begin{eqnarray}
\frac{\partial_r a}{a} &=& \frac{1-a^2}{2r} +\nonumber\\
	&&\frac{1}{4}\kappa_0 r
	\left[\omega^2 \phi_{0}^{2}\frac{a^2}{\alpha^2}
	+(\partial_r \phi_0)^{2} + 
	a^2 \phi_{0}^{2} (m^2 
	+ \frac{1}{2}\lambda \phi_{0}^{2})
	\right],\nonumber\\
\frac{\partial_r \alpha}{\alpha} &=&
	\frac{a^2-1}{r} + 
	\frac{\partial_r a}{a} -
	\frac{1}{2}\kappa_0 r a^2\phi_{0}^{2}(m^2 
	+\frac{1}{2}\lambda\phi_{0}^{2}),\nonumber\\
\partial_{rr}\phi_0  &+& \partial_r \phi_0  \left( \frac{2}{r} + 
	\frac{\partial_r \alpha}{\alpha} - \frac{\partial_r a}{a}\right) 
	+ \omega^2 \phi_0 \frac{a^2}{\alpha^2} \nonumber\\
	&-& a^2 (m^2 + \lambda \phi_{0}^{2}) \phi_0 
	=0.\label{sphericalekgc-sc}
\end{eqnarray}

\noindent This system is a set of 
coupled ordinary differential equations to be solved under the 
conditions of spatial flatness at the origin $a(0)=1$, $\phi_0(0)$ 
finite and $\partial_r \phi_0(0)=0$ in 
order to guarantee regularity and spatial flatness at the origin, and 
$\phi_0(\infty)=\phi^{\prime}_{0}(\infty)=0$ in order to ensure asymptotic 
flatness at infinity as described in 
\cite{ruffini,SeidelSuen1990,Balakrishna1998,gleiser,scott}; these 
conditions reduce the system 
(\ref{sphericalekgc-sc}) to an 
eigenvalue problem for $\omega$, that is, for every central value of 
$\phi_0$ there is a unique $\omega$ with which the boundary conditions
are satisfied. Equations  
(\ref{sphericalekgc-sc})
have several coefficients that can be removed with a rescaling. In order to do so,
the following transformation works: 
$\tilde{\phi}_0 = \sqrt{\frac{\kappa_0}{2}} \phi_0$, 
$\tilde{r} = mr$, 
$\tilde{t} = \omega t$, 
$\tilde{\alpha} = \frac{m}{\omega}\alpha$ and 
$\Lambda = \frac{2\lambda}{\kappa_0 m^2}$. The result is that the physical
constants vanish from the equations, the radial coordinate has
units of $m$ and  time has units of $\omega$. After substituting this transformation and removing the tildes from
everywhere, the resulting EKG system of equations reads:

\begin{eqnarray}
\frac{\partial_r a}{a} &=& \frac{1-a^2}{2r} +\nonumber\\
	&&\frac{1}{2} r
	\left[\phi_{0}^{2}\frac{a^2}{\alpha^2}
	+(\partial_r \phi_0)^{2} + 
	a^2 (\phi_{0}^{2}
	+ \frac{1}{2}\Lambda \phi_{0}^{4})
	\right],\nonumber\\\
\frac{\partial_r \alpha}{\alpha} &=&
	\frac{a^2-1}{r} + 
	\frac{\partial_r a}{a} -
	r a^2\phi_{0}^{2}(1 
	+\frac{1}{2}\Lambda\phi_{0}^{2}),\nonumber\\
\partial_{rr}\phi_0  &+& \partial_r \phi_0  
	\left( 
		\frac{2}{r} + 
		\frac{\partial_r \alpha}{\alpha} - \frac{\partial_r a}{a}
	\right) 
	+ \phi_0 \frac{a^2}{\alpha^2} \nonumber\\
	&-& a^2 (1 + \Lambda \phi_{0}^{2}) \phi_0 
	=0.
\label{sphericalekgc-sc-rescaled}
\end{eqnarray}

\noindent These equations are solved using finite differences with an ordinary 
integrator (adaptive step-size fourth order Runge-Kutta algorithm in the present case) and a shooting routine
that bisects the value of $\omega$.


\begin{figure}[htp]
\includegraphics[width=8cm]{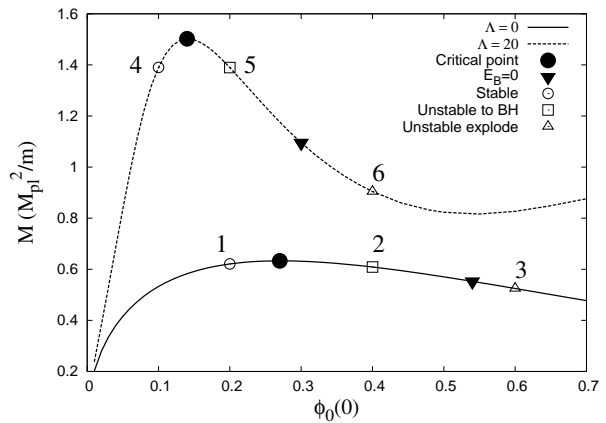}
\caption{\label{fig:equilibrium} Sequences of equilibrium 
configurations for two values of $\Lambda$ are shown as a function 
of the central value of the scalar field $\phi_0(0)$; each point of the curves 
corresponds to a solution of the eigenvalue problem and represents a boson star 
configuration. The filled circles indicate the critical solution that 
separates the stable from the unstable branch. Those
configurations to the left of the maxima represent stable
configurations. The inverted triangles 
indicate the point at which the binding energy is zero. Those 
configurations between the filled circles and the inverted triangles (along each 
sequence) collapse into black holes as a response to a perturbation. 
Configurations to the right of the inverted triangles disperse away.}
\end{figure}

The solutions of (\ref{sphericalekgc-sc-rescaled}) define sequences of 
equilibrium configurations like those shown in Fig. 
\ref{fig:equilibrium}. Each point in the curves corresponds to a
boson star solution. In each of the curves two 
important points for each value of $\Lambda$ are marked: i) the critical 
point -marked with a filled circle- indicating the threshold between the 
stable and unstable branches of each sequence, that is, configurations 
to the left of this point are stable and those to the right are unstable 
as found through the analysis of perturbations \cite{gleiser,scott}, 
catastrophe theory \cite{Catastrophe} and full non-linear evolution of 
the equilibrium solutions 
\cite{SeidelSuen1990,Balakrishna1998,Guzman2004} 
and ii) the point at which the binding energy $E_B = M-Nm = 0$ marked 
with an inverted filled triangle (see \cite{gleiser} for this convention of the binding 
energy), where $N=\int 
j^0 d^3 x = \int 
\frac{i}{2}\sqrt{-g}g^{\mu\nu}[\phi^{*}\partial_{\nu}\phi - \phi
\partial_{\nu}\phi^{*}]d^3 x$ is the number of particles; that is,
the conserved quantity due to the invariance under the global $U(1)$
group of the Lagrangian density (\ref{eq:lagrangian}). $M =
(1-1/a^2)r/2$ evaluated at the outermost point of the numerical domain 
is the Misner-Sharp mass; the configurations between the instability 
threshold and the zero binding energy point have negative binding energy
($E_B<0$) and collapse into black holes whereas those to the right 
have positive binding energy and are expected to disperse away. Stable configurations, obviously have negative binding energy.
The non-filled circles, triangles and squares 
pointed out in the plot  correspond to the six 
configurations chosen as special cases to be explored here. 


\section{The evolution of boson stars}
\label{sec:evolution}

In order to verify the fate of Boson Stars one has to evolve them. For this it is necessary to relax the condition of stationarity of space-time and allow the scalar field and space-time metric to evolve. Then Boson Star configurations constructed so far, both scalar field profile and metric functions, will work only as initial conditions for an initial value problem associated to the EKG equations.

For the construction of the evolution equations we split  the scalar field into its real and imaginary parts
$\phi = \phi_1 + i \phi_2$. In this way, the Klein Gordon 
equation becomes two equations:

\begin{equation}
\left(
\Box - \frac{dV}{d|\phi|^2}
\right)\phi_1 = 0, ~~~~
\left(
\Box - \frac{dV}{d|\phi|^2}
\right)\phi_2 = 0.
\label{eq:bs2kg}
\end{equation}

\noindent Another major point is that as time-dependence of the
space-time is going to be allowed, the space-time metric has to be 
relaxed to:

\begin{equation}
ds^2=-\alpha(r,t)^2dt^2 + a(r,t)^2dr^2 + r^2 d\Omega^2, 
\label{eq:spherical_metric_t}
\end{equation}

\noindent where time dependence of $\alpha$ and $a$ has been allowed, but
the gauge is such that the radial coordinate has been kept the same. Using this
new line element for the space-time, it is convenient to define 
first order variables for the scalar field: $\pi_i = 
\frac{a}{\alpha}\partial_t \phi_i$ and $\psi_i = \partial_r\phi_i$, for 
each $i=1,2$. With these new variables and  metric 
(\ref{eq:spherical_metric_t}) the KG equations are translated into 
the following set of first order PDEs

\begin{eqnarray}
\partial_t \phi_1 &=& \frac{\alpha}{a}\pi_1, \nonumber\\
\partial_t \phi_2 &=& \frac{\alpha}{a}\pi_2, \nonumber\\
\partial_t \psi_1 &=& \partial_r\left(
	\frac{\alpha}{a} \pi_1 \right),\nonumber\\
\partial_t \psi_2 &=& \partial_r\left(
	\frac{\alpha}{a} \pi_2 \right),\label{eq:bsevolutonjoder}\\
\partial_t \pi_1 &=& \frac{1}{r^2} \partial_r\left(
	r^2\frac{\alpha}{a}\psi_1\right) - a\alpha
	\frac{dV}{d|\phi|^2}\phi_1,\nonumber\\
\partial_t \pi_2 &=& \frac{1}{r^2} \partial_r\left(
	r^2\frac{\alpha}{a}\psi_2\right) - a\alpha
	\frac{dV}{d|\phi|^2}\phi_2.\nonumber
\end{eqnarray}

\noindent On the other hands, in terms of the first order variables and the line element 
(\ref{eq:spherical_metric_t}), Einstein's equations read:

\begin{eqnarray}
\frac{\partial_r a}{a} &=& \frac{1-a^2}{2r} + 
	\frac{r}{2}[\pi_1^2 + \pi_2^2
	+ \psi_1^2 + \psi_2^2 + a^2 V],\label{eq:bshamc}\\
\frac{\partial_r \alpha}{\alpha} &=& \frac{a^2-1}{r} + \frac{a'}{a} -
	r a^2 V,\label{eq:bsslic}\\
\partial_t a &=& \alpha r[\psi_1 \pi_1 
	+ \psi_2 \pi_2].
\label{eq:bsconstraints}
\end{eqnarray}

\noindent These equations correspond to the
Hamiltonian constraint, the slicing condition and the 
momentum constraint. 
Clearly, this set of three equations is over-determined, and it is 
necessary to choose two of these three equations as the set to be solved.

The  system of equations to be solved is (\ref{eq:bsevolutonjoder}) together with (\ref{eq:bsconstraints}), which constitutes a constrained evolution system. Among the Einstein's equations we choose to solve (\ref{eq:bshamc}) and (\ref{eq:bsslic}) and only monitor the validity of (\ref{eq:bsconstraints}) during the evolution.

The solution is as follows. Equations (\ref{eq:bsevolutonjoder}) are solved using the method of lines with second order accurate stencils and a third order Runge-Kutta time integrator. Constraint equations (\ref{eq:bshamc}) and (\ref{eq:bsslic}) are solved using a fourth order Runge-Kutta ODE solver at all intermediate steps required during the integration of (\ref{eq:bsevolutonjoder}).

The boundary condition for the scalar field is that of 
an an outgoing spherical wave on a Schwarzschild background, which in its
differential equation form reads:

\begin{equation}
\partial_r \pi_i + \partial_t \pi_i + \pi_i/r =0, ~~~~~
\psi_i = -\pi_i - \phi_i/r,
\label{eq:bcpsi}
\end{equation}

\noindent for each $i=1,2$. That is, the real and 
imaginary parts of the scalar field are considered to behave as outgoing 
spherical waves separately. This is a quite simple boundary condition that
certainly can be improved, the only requirement demanded in the present
analysis is that the boundary conditions allow second order convergence 
after various crossing times. These conditions were implemented using a second order  accurate upwind stencils.

\section{Preparing the states}
\label{sec:initialdata}

\begin{table}[h]
\begin{tabular}{cllllll}\hline
Label	&	$M$		&	$M_{pert}$	&	\%	&	$r_{95}$	&	$r_0$	&	A\\\hline\hline
C1	&	0.620882	&			&	0.	&	7.66	&		&		\\
C2	&	0.608758	&	0.60937		&	0.1	&	4.84	&	6.	&	0.0008	\\
C3	&	0.5248		&	0.52537		&	0.1086	&	3.68	&	5.	&	0.0012	\\
C4	&	1.390156	&			&	0.	&	12.27	&		&		\\
C5	&	1.389544	&	1.39101		&	0.105	&	7.62	&	9.	&	0.0008	\\
C6	&	0.9043		&	0.905234	&	0.103	&	5.66	&	7.	&	0.0008	\\\hline
\end{tabular}
\caption{\label{tab:tabla} Configurations used to present the three different
types of behavior. The properties of the Gaussian shell acting as perturbation
and the magnitude in percent of the perturbation are mentioned. The perturbation profile
is $A\ exp(-(r-r_0)^2/0.1)$. $M$ is the mass of the unperturbed Boson Star, $M_{pert}$ is the mass
of the perturbed Star, \% indicates the percent of mass the pertirbation contributes with,
$r_{95}$ is the radius of the Star containing 95 per cent of the total mass of the star.}
\end{table}

\begin{figure*}[htp]
\includegraphics[width=8cm]{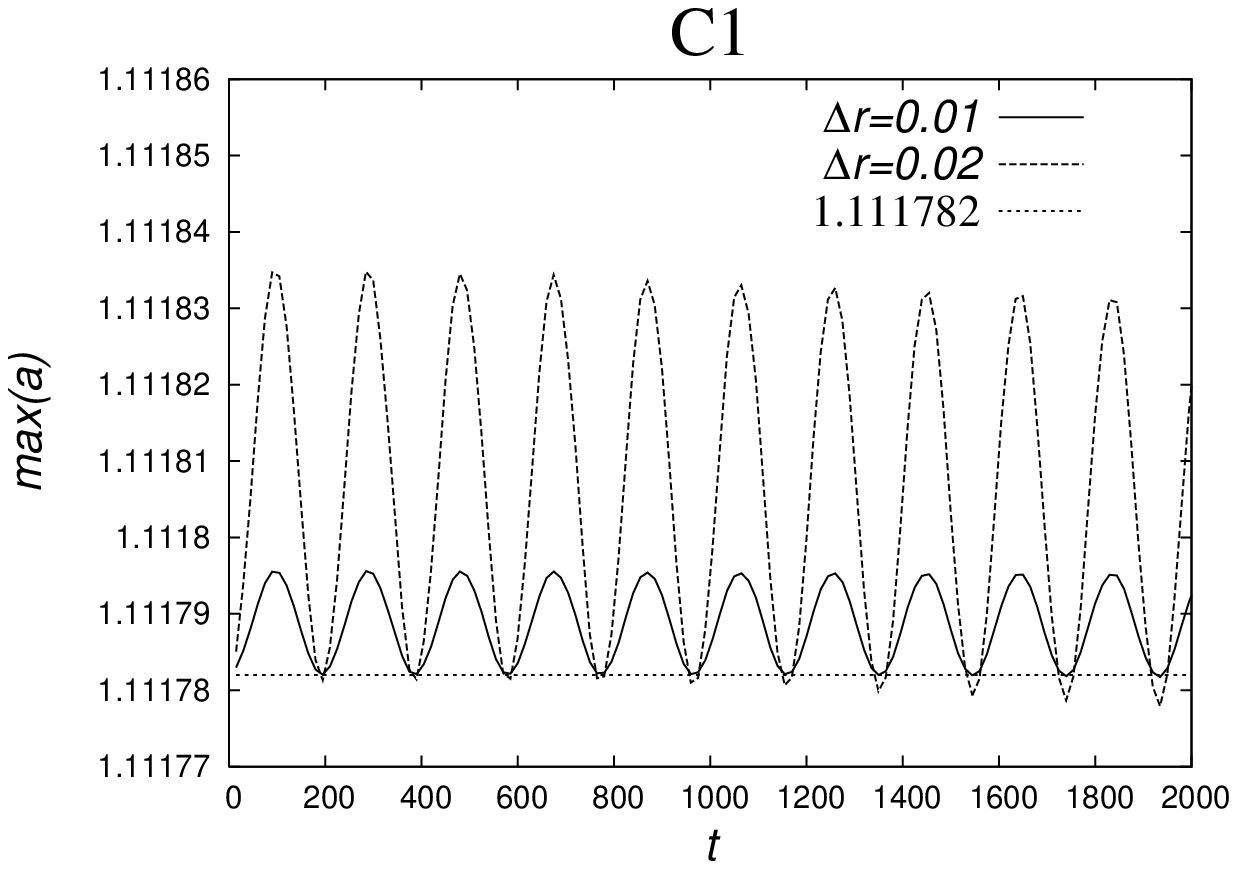}
\includegraphics[width=8cm]{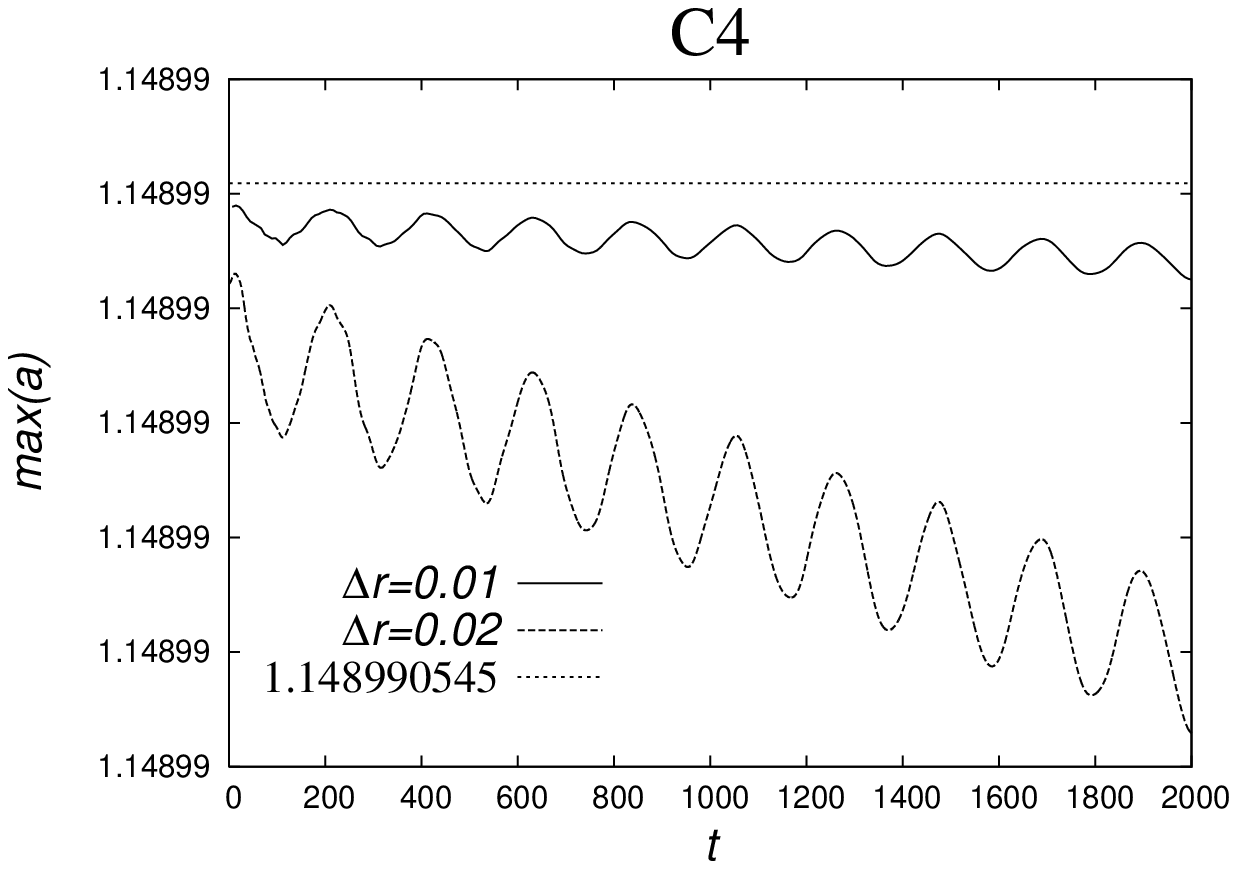}
\includegraphics[width=8cm]{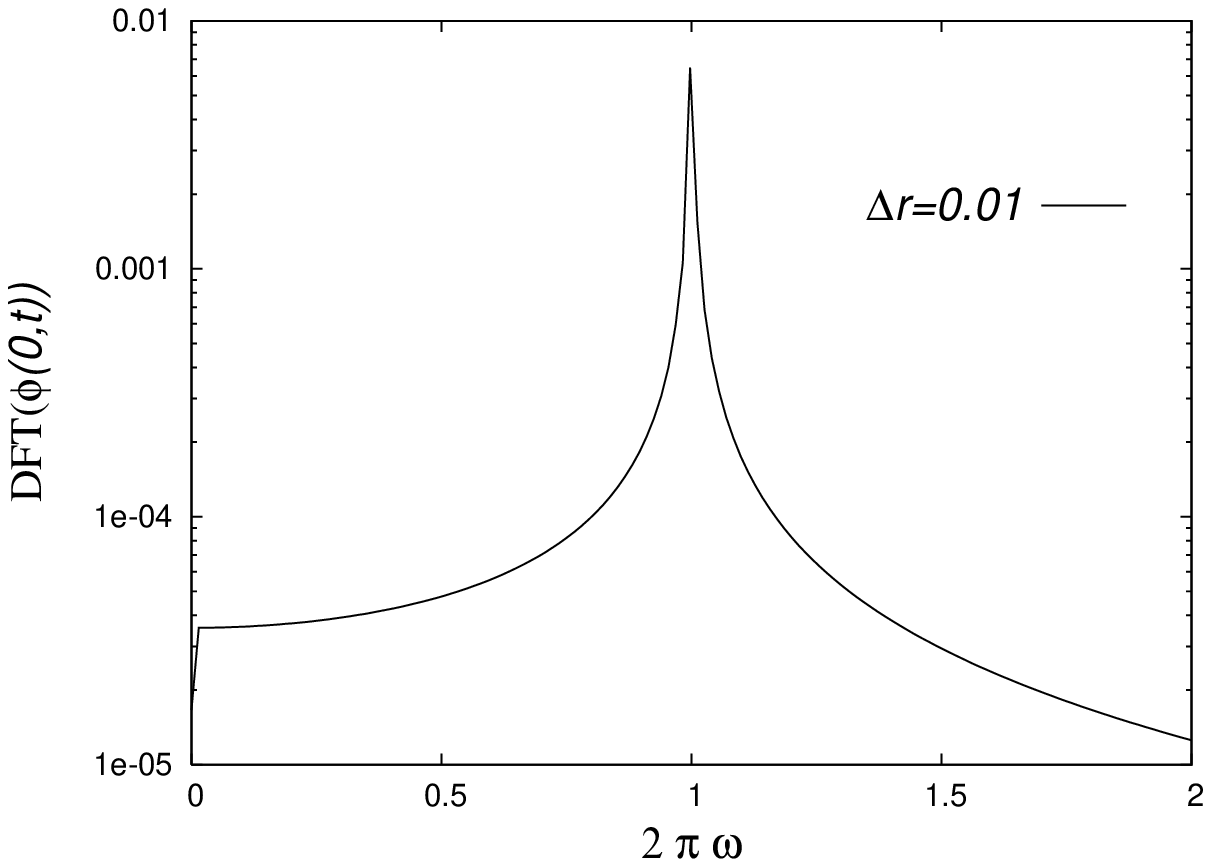}
\includegraphics[width=8cm]{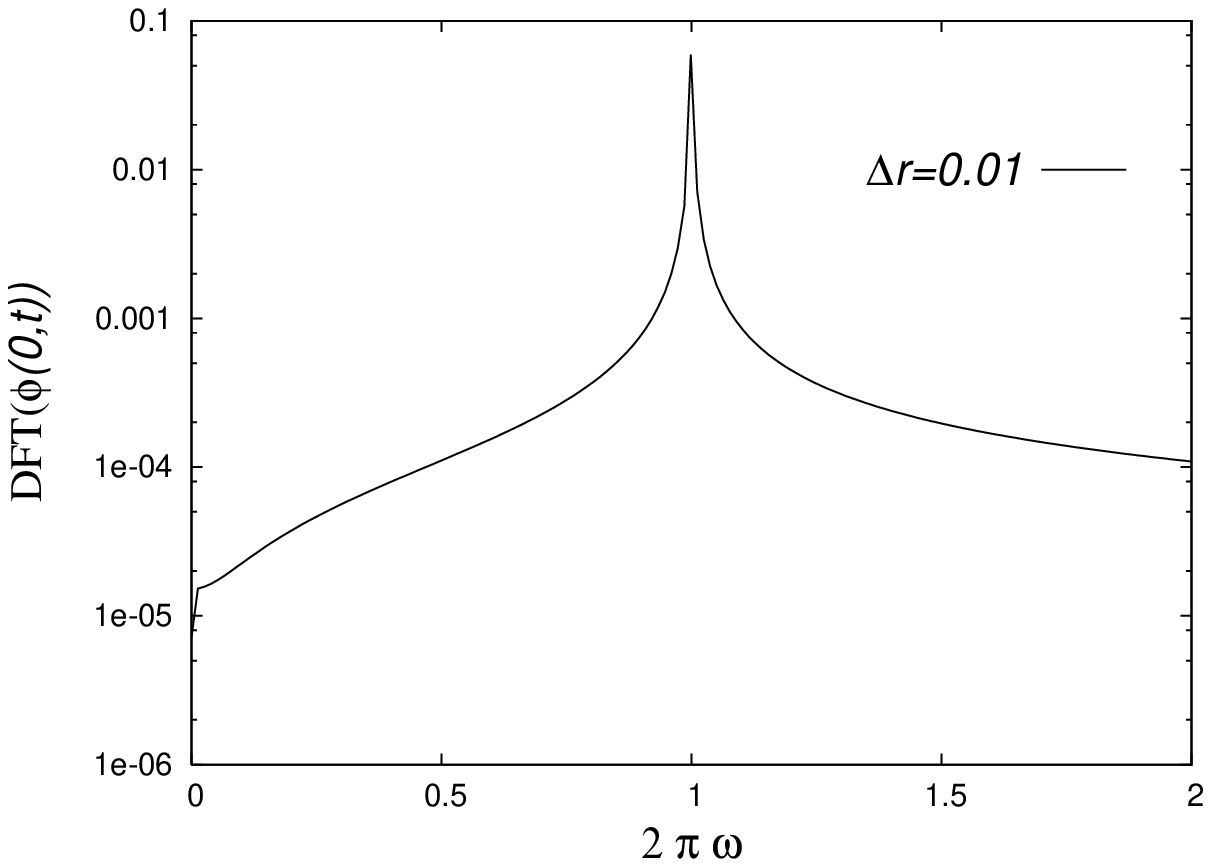}
\includegraphics[width=8cm]{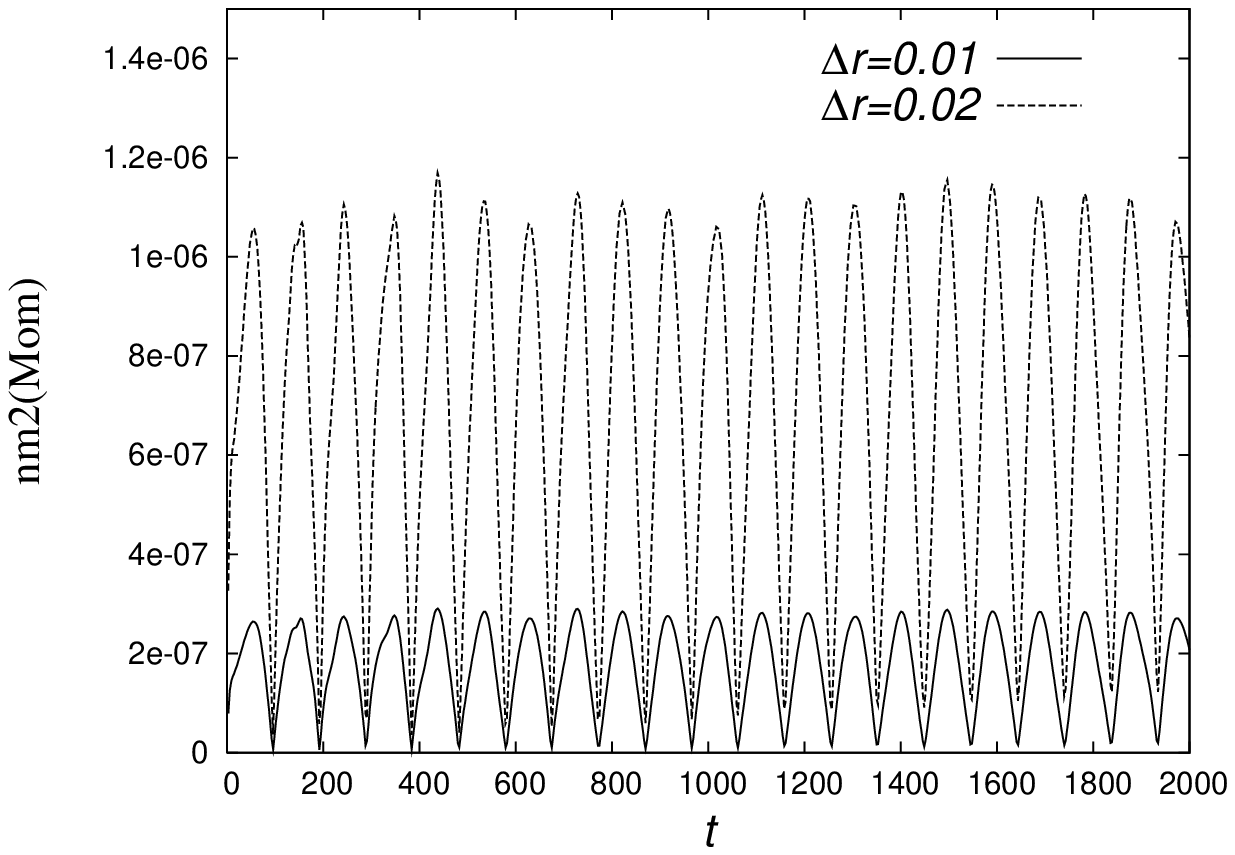}
\includegraphics[width=8cm]{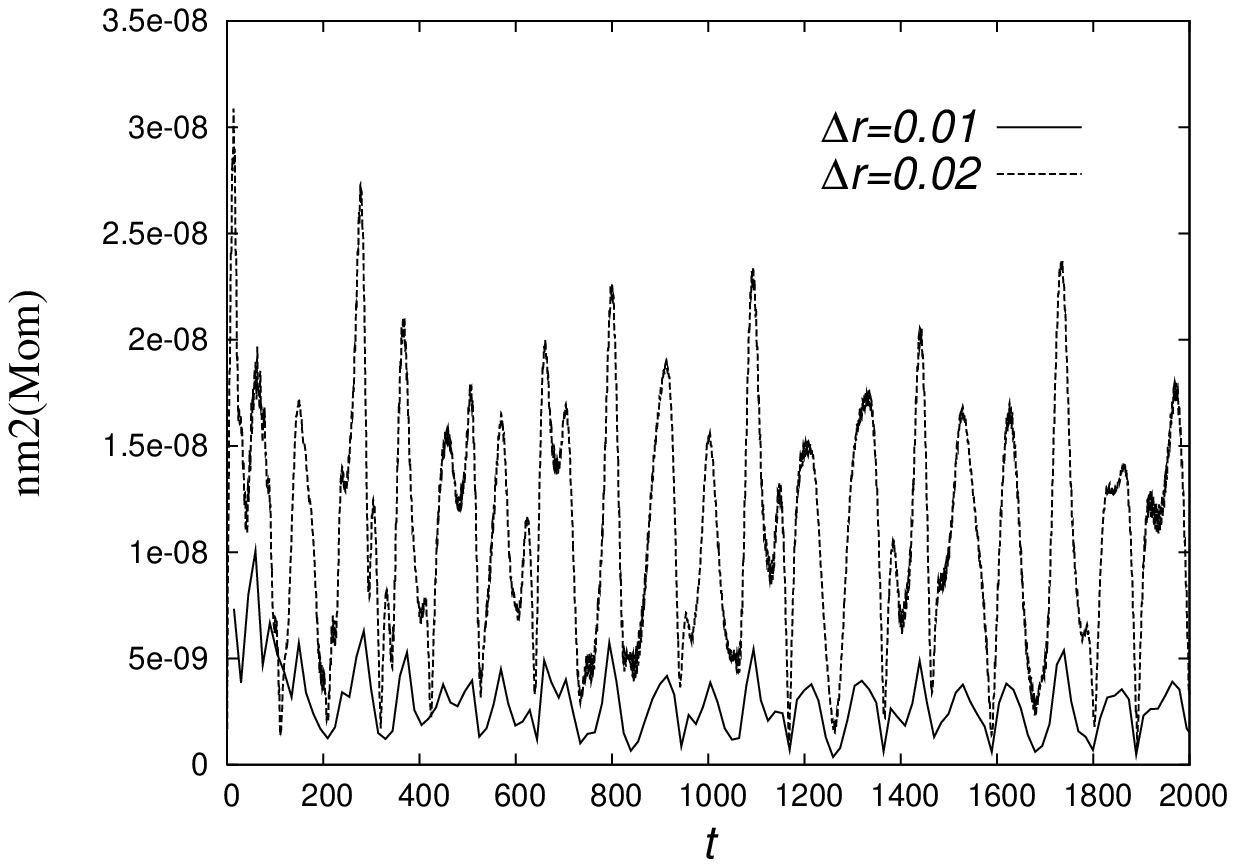}
\caption{\label{fig:stable_amax} (top) Convergence of $a_{max}$ for configurations 1 and 4.
The constant line indicates the value of $a_{max}$ at initial time, which is assumed to
be the value it should keep during the evolution. Second order convergence to such
value during the evolution is a good indicator that the evolution is being carried 
out properly. (Middle) Fourier Transform of the central
value of the field for the same configurations. The peak shows up at
$\omega = \frac{1}{2\pi}$. This test indicates that the scalar field
is oscillating with the correct frequency. (Bottom) Convergence of the 
$L_2$ norm of the momentum constraint;
the aim of these plots is to show that the
momentum constraint is satisfied in the continuum limit. The numerical parameters are as indicated in 
the plots: two resolutions were used for the convergence test.}
\end{figure*}

\begin{figure*}[htp]
\includegraphics[width=8cm]{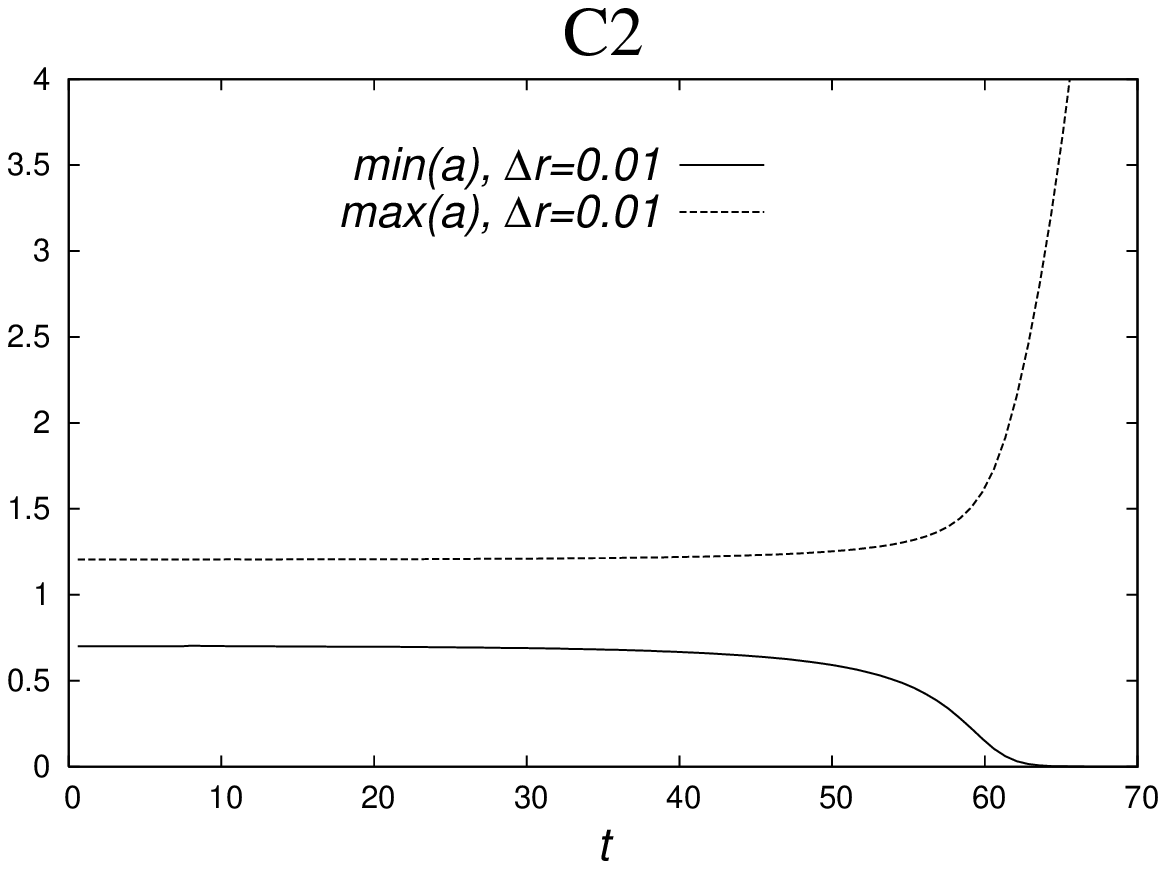}
\includegraphics[width=8cm]{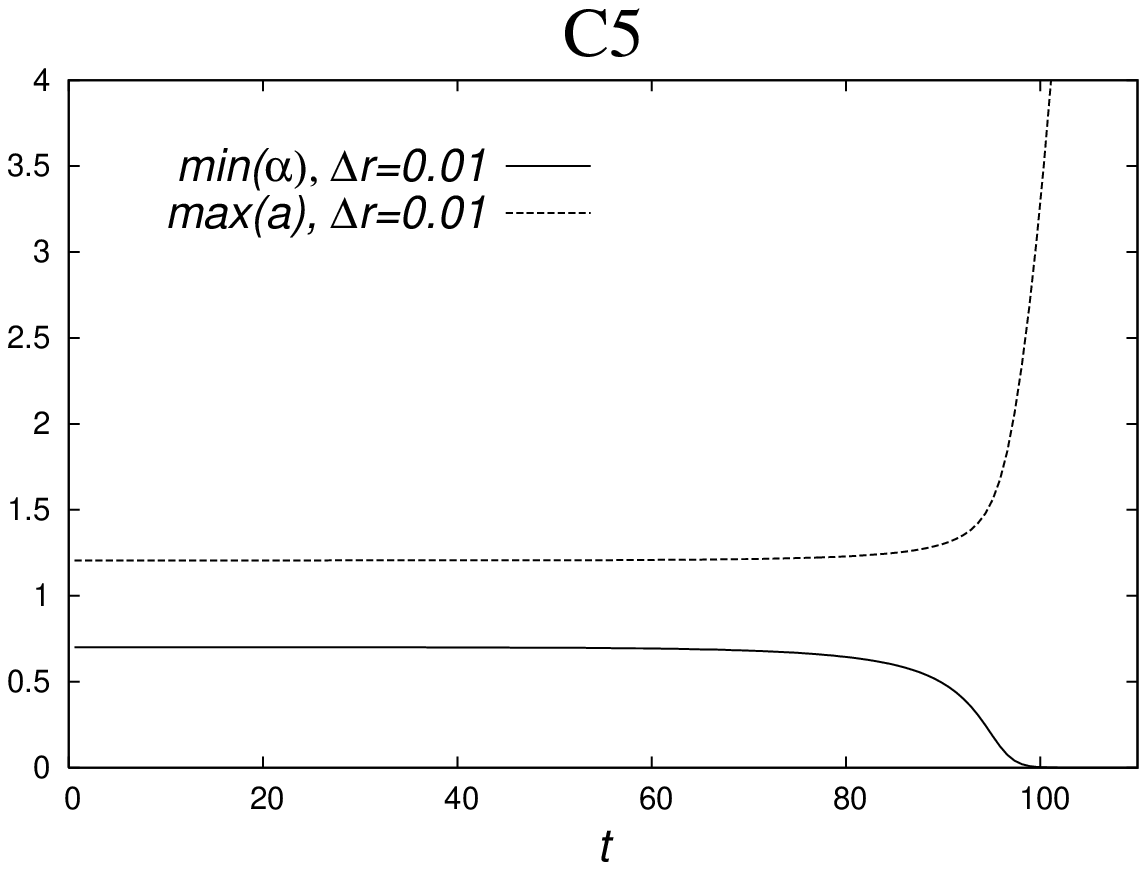}
\includegraphics[width=8cm]{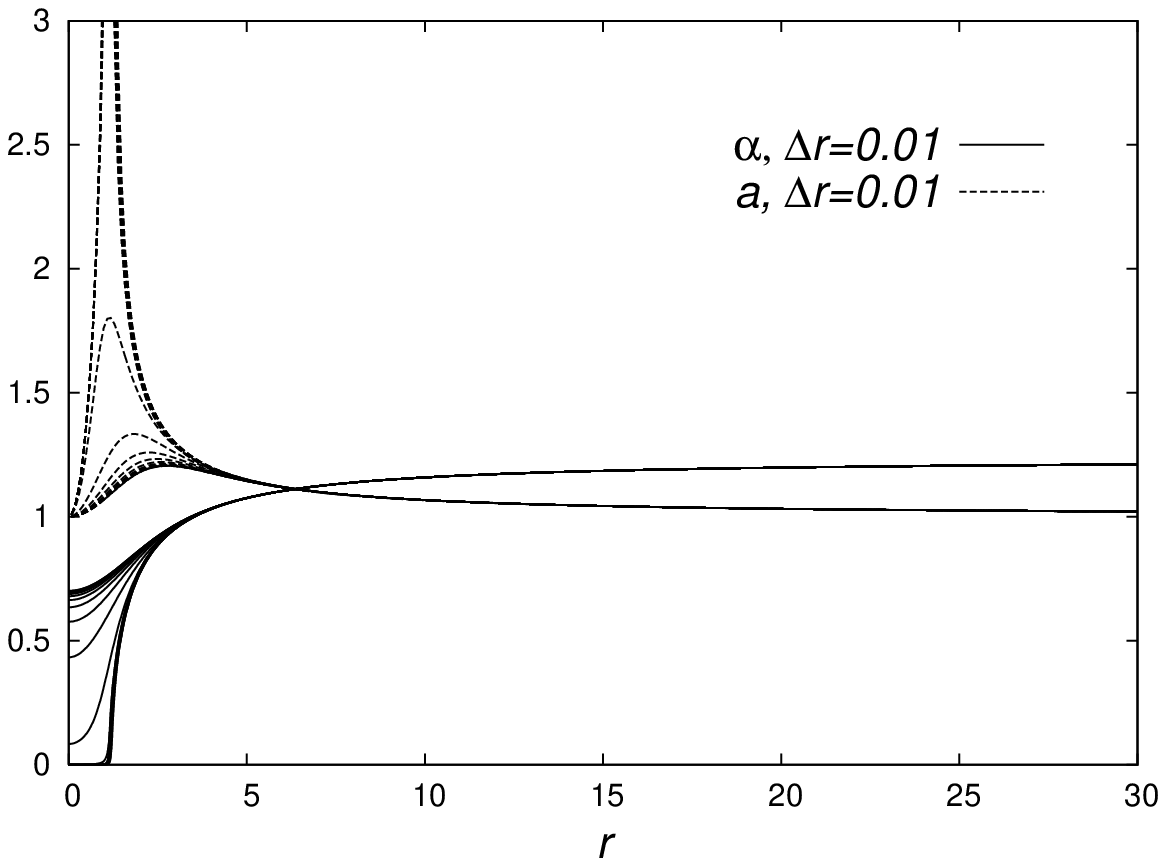}
\includegraphics[width=8cm]{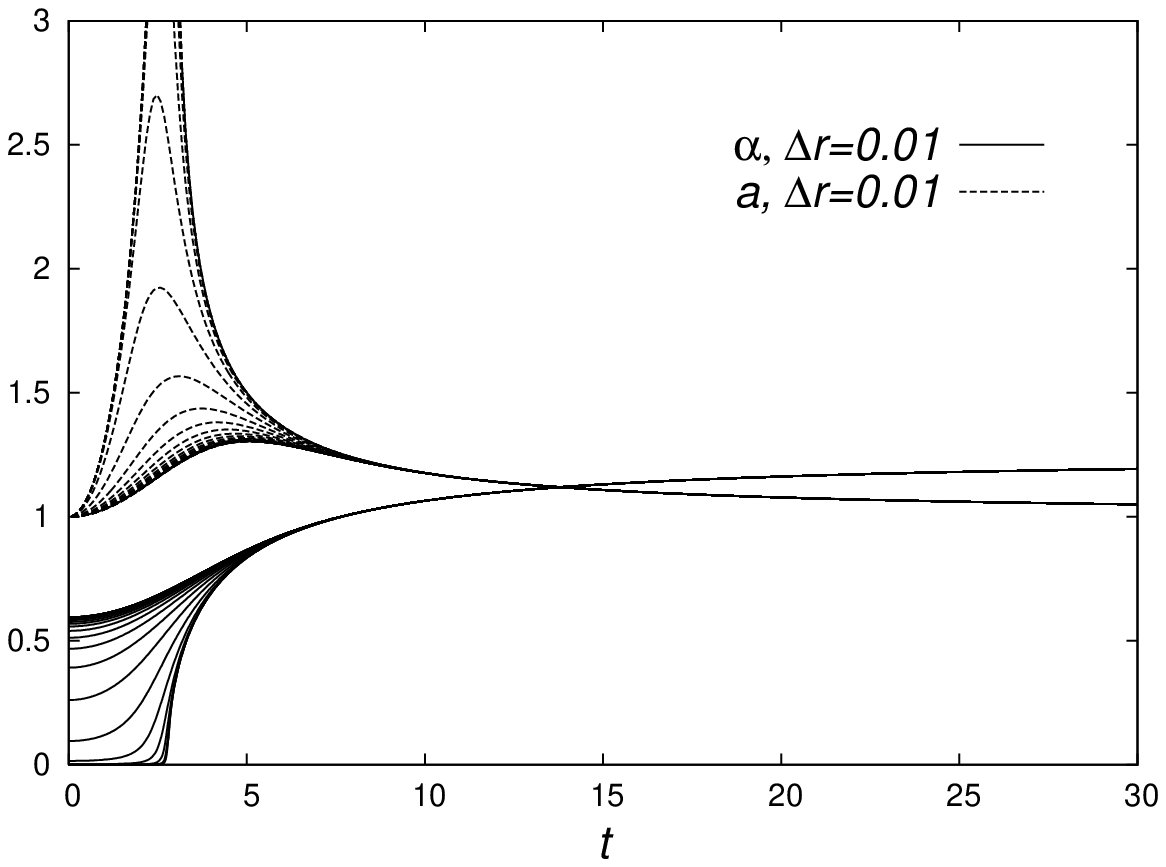}
\caption{\label{fig:collapsing} (Top) Maximum of
the metric function $a$ and minimum of $\alpha$. What is shown
is the collapse of the lapse and the divergence of $a$, indicating 
the formation of a horizon in the coordinates that are being used.
(Bottom) Snapshots of $a$ and $\alpha$ that show the process of collapse.
The resolution used is the one shown in the plots.}
\end{figure*}

\begin{figure*}[htp]
\includegraphics[width=8cm]{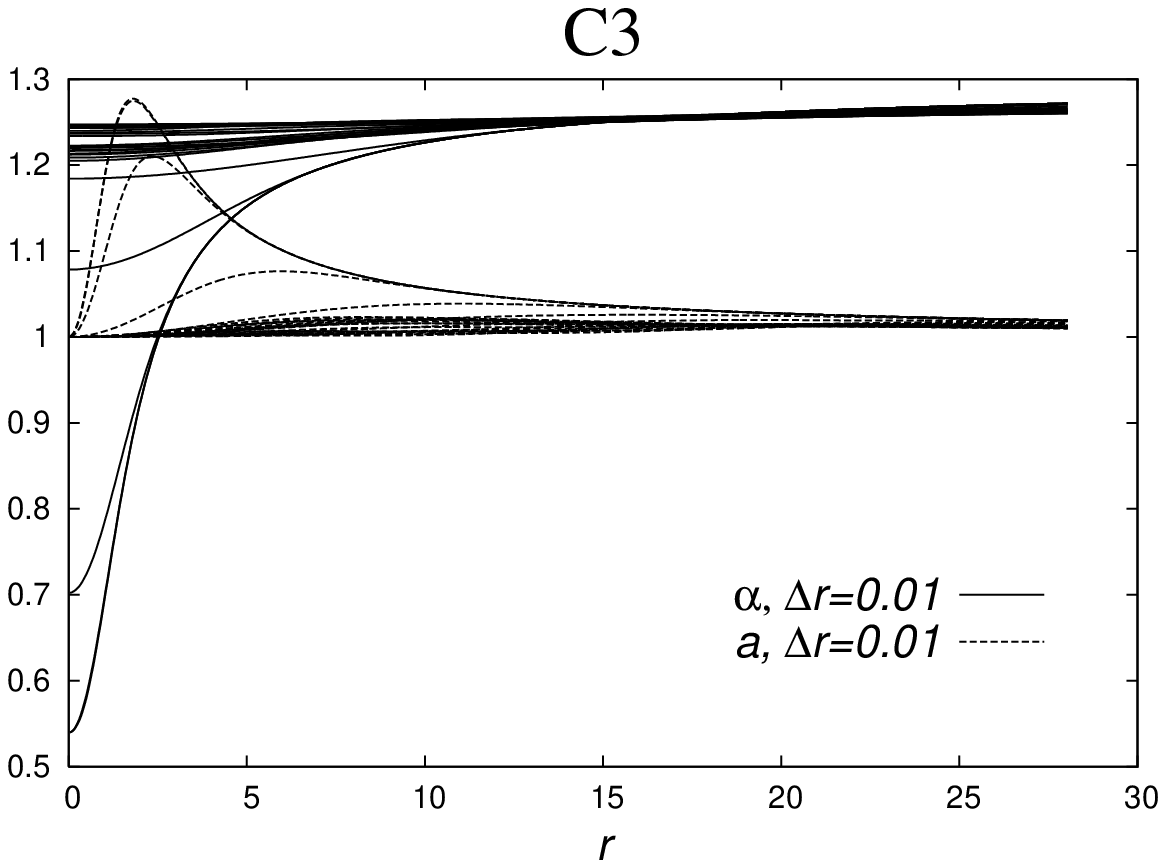}
\includegraphics[width=8cm]{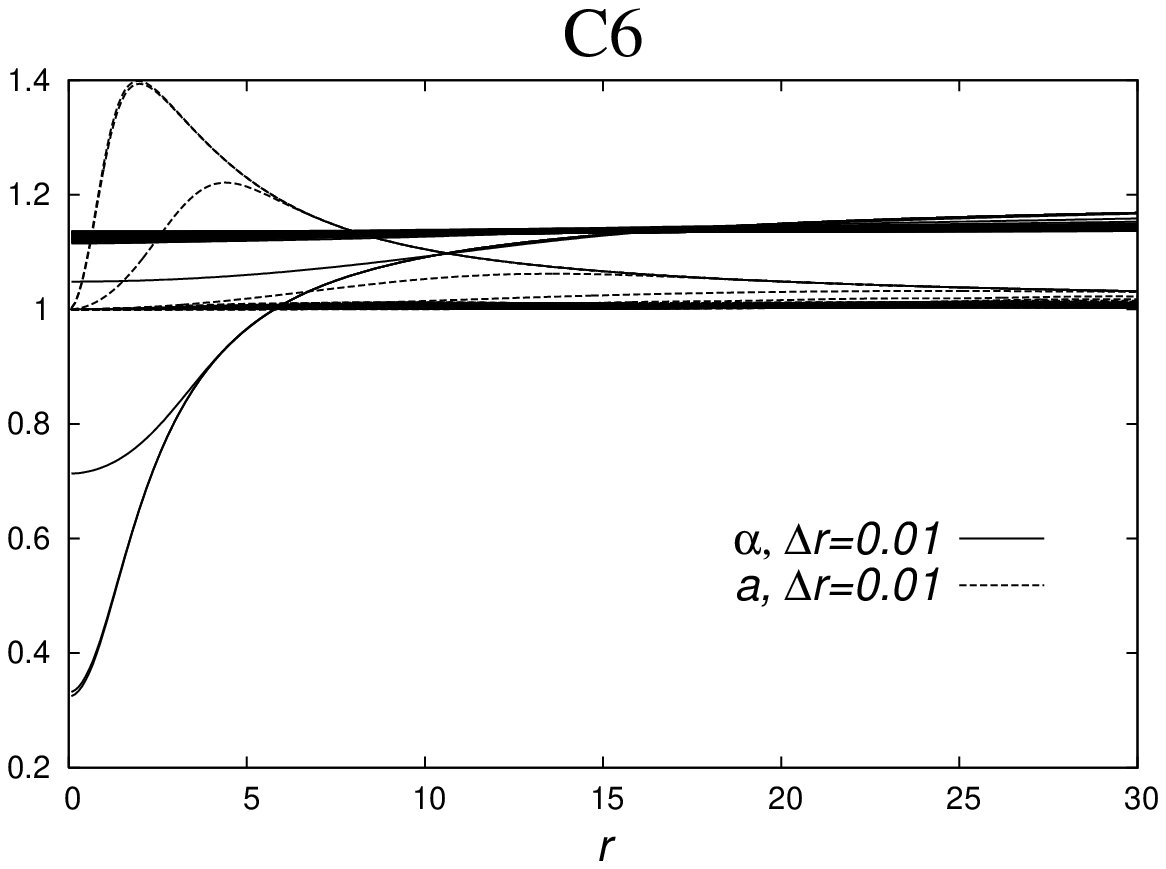}
\includegraphics[width=8cm]{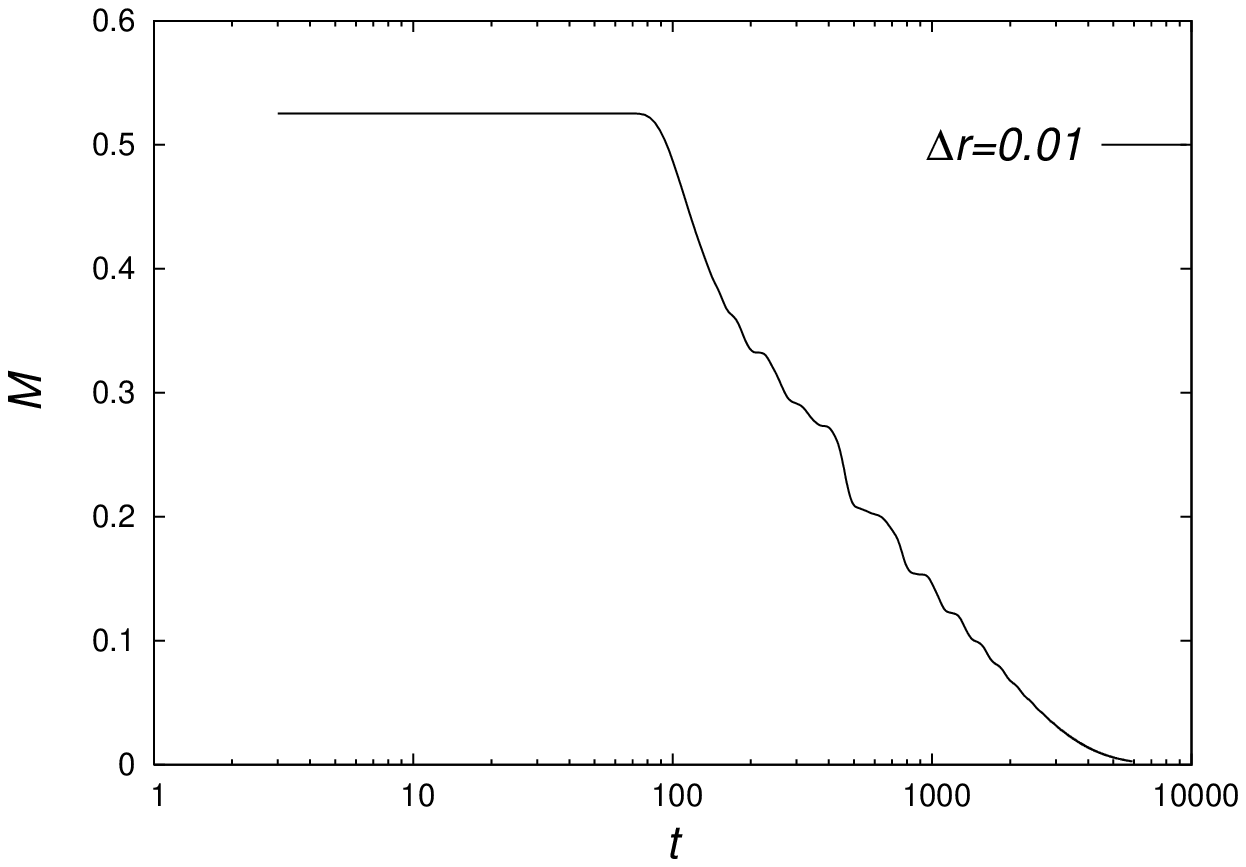}
\includegraphics[width=8cm]{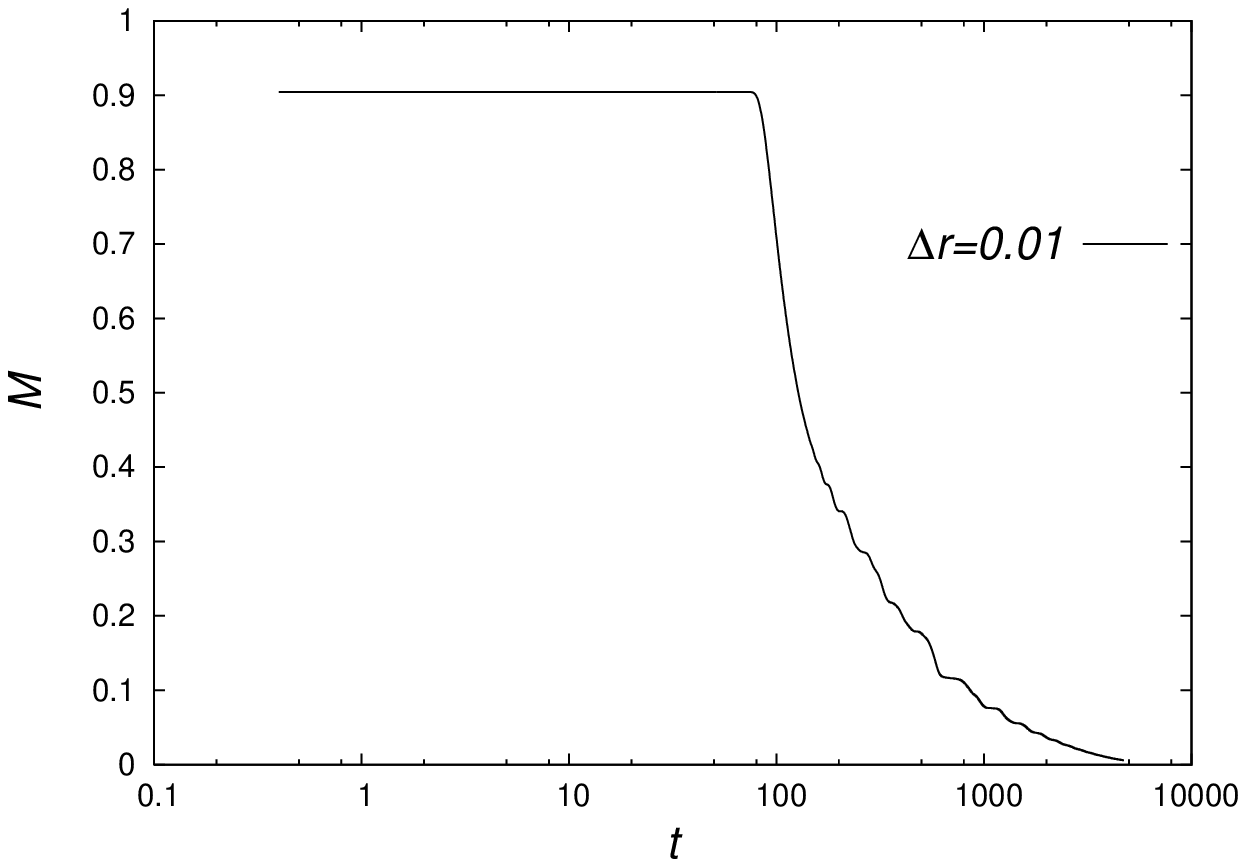}
\includegraphics[width=8cm]{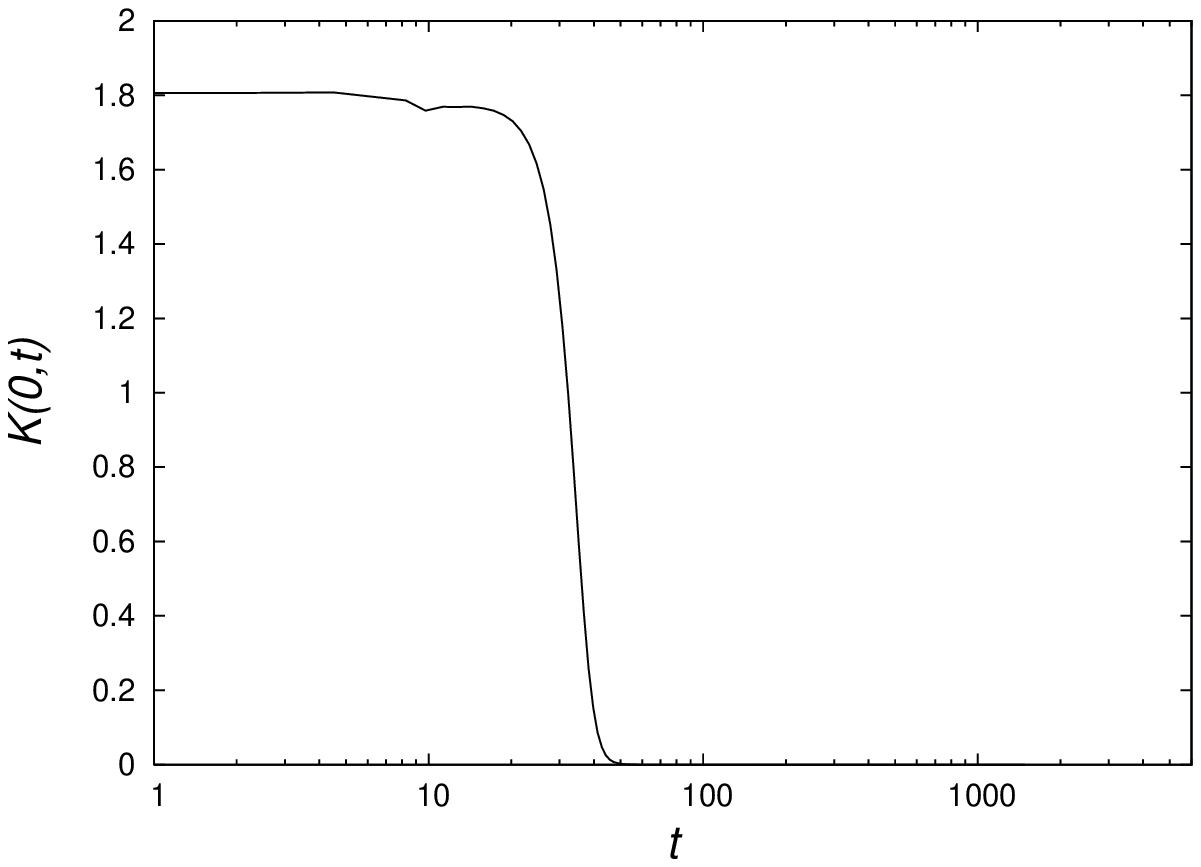}
\includegraphics[width=8cm]{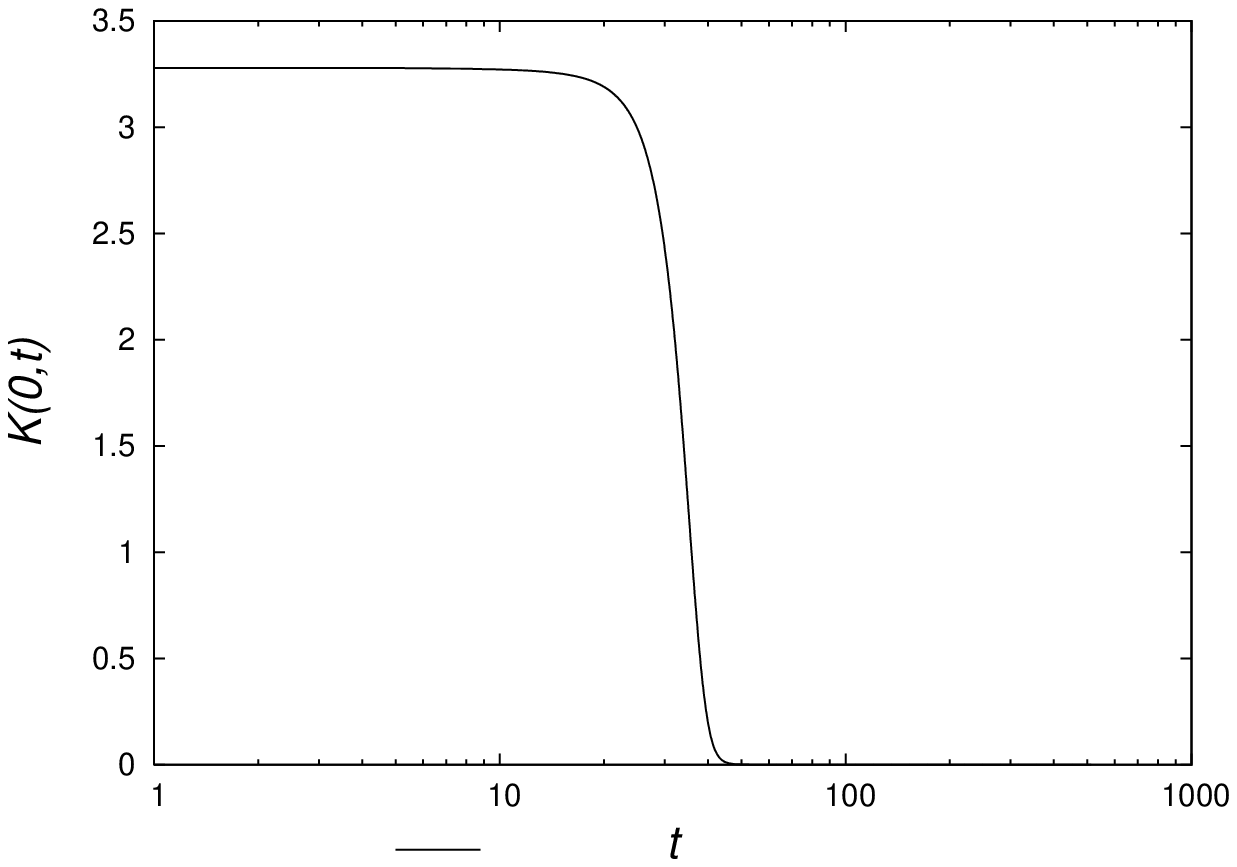}
\caption{\label{fig:exploding} (Top) Snapshots of the metric functions
for configurations C3 and C6, illustrating that the metric evolves toward 
the flat metric. (Middle) Mass vs time of the system, which shows that
the configurations are releasing all its mass toward infinity.
(Bottom) Central value of the Kretchmann scalar, which tends to zero 
with the pass of time. The resolution used is indicated 
in the plots.}
\end{figure*}

In order to illustrate the three types of fate for Boson stars, we choose three
configurations for each of the two values of the self-interaction 
$\Lambda=0,20$, that clearly indicate the expected behavior.
According to the labels in Fig. \ref{fig:equilibrium}, 
the theory predicts that configurations 1 and 4 should remain
stable, configurations 2 and 5 should collapse to black holes and configurations
3 and 6 should disperse away. 

In Table \ref{tab:tabla} we summarize the 
properties of the configurations chosen. The selected stable configurations were not 
perturbed so that it was possible to track the convergence toward a time-independence
geometry, whereas we used a gaussian shell to trigger the instability of  
unstable configurations.


\section{Stable configurations}
\label{sec:test1}

The test consists in showing the validity of the hypotheses 
used in the construction of BS solutions (harmonic time dependence of 
the scalar field and time independence of the metric functions). 
The configurations proposed for the test are the configurations marked 
with unfilled circles and numbers 1 and 4 in Fig. \ref{fig:equilibrium}. 

In the top of Fig. \ref{fig:stable_amax} the maximum of the metric function $a$ is
shown.  Due to the discretization error, the configurations are perturbed
permanently, and therefore the metric is oscillating in time. The point
is that one has to show  the amplitude of the oscillations converges to zero in the continuum limit,
which is equivalent to have convergence of $a_{max}$ to the value it should
keep during the evolution, which we consider to be the value of $a_{max}$
calculated for the equilibrium configuration at initial time. Since the stencils of the finite differences used to solve the evolution system are second order accurate, the departure from the equilibrium configuration should reduce by a factor of four when doubling the resolution, which is exactly what is shown at the top of Fig. \ref{fig:stable_amax}, because the amplitude of oscillations is four times bigger when using resolution $\Delta r=0.02$ than when using $\Delta r=0.01$. 

Another test of the correct evolution consists in showing that the scalar field is truly 
oscillating meanwhile with the correct frequency. In the middle panel
of Fig. \ref{fig:stable_amax} the Fourier Transform of the central value
of $\phi_1$ is shown. The result indicates that the frequency of oscillation 
of the scalar field corresponds to the eigenvalue calculated when solving the 
initial value problem (remember the time was rescaled to be 
$\tilde{t} = \omega t$). In the units used here, where $t \rightarrow 
\omega t$, the frequency is $\omega = 1/2\pi$).

Finally let us discuss the momentum constraint. Since the stencils of the finite differences used to solve the evolution system are second order accurate, the momentum constraint (\ref{eq:bsconstraints}) is not satisfied exactly, instead $\partial_t a - \alpha r[\psi_1 \pi_1 + \psi_2 \pi_2] ={\cal O}(\Delta r^2)$, therefore the violation of this constraint should reduce quadratically with resolution. For the verification of convergence it suffices to check whether a norm of the violation, which is actually defined at each time step over the spatial domain, converges.
In the bottom panel of Fig. \ref{fig:stable_amax} we show the $L_2$ norm of the violation as a function of time, using  resolutions $\Delta r=0.01,0.02$ and the fact that the violation using the coarse resolution is four times bigger than with the fine resolution, shows the second order convergence of the violation to zero in the continuum limit.

No explicit perturbations are applied in the stable case. Instead, 
measuring the correct frequency of oscillation of the scalar field 
determines whether or not the evolution code is working properly or 
not. Applying explicit perturbations to the particle number of these 
stable configurations would imply a more dynamical behavior, and 
the amplitude decay of the maximum of the metric function $a$ could be studied as done
in \cite{Balakrishna1998}.


\section{Bounded unstable configurations}
\label{sec:test2}

For this second scenario we choose the adequate configurations
2 and 5, which have a negative binding energy. 
In principle, the discretization error would provide
a strong enough perturbation to collapse the configuration. However,
in order to have a quicker collapse we use a small perturbation
consisting of the addition of a  
Gaussian shell to the real part of the scalar field, with
properties shown in Table \ref{tab:tabla}. Then the
equations for $\alpha$ and $a$ are resolved before starting 
the evolution. The amplitude of the Gaussian shell is positive and the number of
particles is increased by a small fraction of its initial value.

The results of the evolution are 
summarized in Fig. \ref{fig:collapsing}, where snapshots of the 
lapse and $a$ are shown; in fact the lapse collapses to zero in a region 
expected to be covered by a horizon and the metric function $a$
starts diverging due to the slice-stretching effect when using normal coordinates \cite{Alcubierre}. 
In the coordinates used, an 
apparent horizon has been formed when the lapse is sufficiently near 
to zero. However, it is simple to use different coordinates allowing one to 
calculate the location, mass and possible oscillations of an apparent 
horizon. The whole process where apparent and event horizons was calculated during the collapse of an unstable BS can be found in \cite{Guzman2004}. 
Up to this point the results in this section correspond to the 
typical results found for spherically symmetric BSs in the canonical 
papers \cite{SeidelSuen1990,Balakrishna1998}.


\section{Unbounded unstable configurations}
\label{sec:test3}

In this case, the binding energy is positive, which means that 
there is no work needed to disassemble the configuration,  
a case that is barely mentioned in BSs studies. A previous case of fissioned BS 
was found in \cite{Guzman2004}, in the context of the full 3D unconstrained evolution of BSs in 3+1 Numerical Relativity.

In Fig. \ref{fig:exploding} the results for configurations 3 and 6 
are shown with three different elements: i) the metric function $a$ and
the lapse $\alpha$ become constant after a while during the evolution,
ii) the mass function $M$ decays to zero during the same time-scale, and 
iii) the central value of the Kretchmann scalar $K$ becomes zero, in order to
have an indication that no singularities are left behind the exploding scalar field. 
This is supported by the fact that the metric at the origin is spatially flat  with a non-zero lapse.

The perturbation consisted in the addition of a tiny fraction of particles again, 
even though, the configurations reacted in a very explosive way. The property assumed to be the responsible for this explosive behavior is the one argued as 
the responsible to work against the gravitational collapse of the bosons, that is, the 
uncertainty principle, as oposed to the degeneracy presure that prevents the gravitational 
collapse of fermionic stars. The implication of the uncertainty 
principle is therefore that the configuration has an excess of kinetic energy to compensate 
the localization of the wave function.


\section{Final remarks}
\label{sec:remarks}

Among the three types of boson stars presented here, only the stable
type has been studied with certain attention, and stable configurations
are proposed either as potential existing objects or as toy models.
Perhaps unstable configurations might impose restrictions on mass and self-interaction of Boson Stars as models that attempt to supplant black holes \cite{diego-acc,Guzman2006,RuedaGuzman2009}, or in the low energy and weak field limit, scalar fields playing the role of -for instance- dark matter \cite{GuzmanUrena2006} might also be restricted.


\appendix
\section{Kretschmann invariant}

We use the central value of the Kretschmann invariant, in order
to easily see the formation of a singularity. The expression for the metric 
(\ref{eq:spherical_metric_t}) is:

\begin{eqnarray}
K &=& -\frac{16 \dot{a}^2 }{r^2 \alpha^2 a^4} 
      + \frac{4 (\alpha^{\prime\prime})^2}{\alpha^2 a^4}
      -\frac{8\alpha^{\prime\prime} \ddot{a}}{\alpha^3 a^3} 
      + \frac{8\alpha^{\prime\prime} \dot{\alpha}\dot{a}}{\alpha^4 a^3}\nonumber\\
  &-& \frac{8\alpha^{\prime\prime} \alpha^{\prime}a^{\prime} }{\alpha^2 a^5}
      +\frac{4(\ddot{a})^2}{\alpha^4 a ^2}
      -\frac{8\ddot{a}\dot{\alpha}\dot{a}}{\alpha^5 a^2}
      +\frac{8 \ddot{a} \alpha^{\prime} a^{\prime} }{\alpha^3 a^4}\nonumber\\
  &+& \frac{4 (\dot{\alpha})^2 (\dot{a})^2 }{\alpha^6 a^2}
      -\frac{8 \dot{\alpha} \dot{a} \alpha^{\prime} a^{\prime}}{\alpha^4 a^4}
      +\frac{4(\alpha^{\prime})^2 (a^{\prime})^2}{\alpha^2 a^6}
      +\frac{8 (\alpha^{\prime})^2}{r^2 \alpha^2 a^4}\nonumber\\
  &+& \frac{8 (a^{\prime})^2}{r^2 a^6}
      +\frac{4}{r^4}
      -\frac{8}{r^4 a^2}
      +\frac{4}{r^4 a^4}.
\nonumber
\end{eqnarray}


\section*{Acknowledgments}

This work is partly supported by projects CIC-UMSNH-4.9, 
CONACyT 79995 and PROMEP-UMICH-CA-22.



\end{document}